\documentclass[pre,aps,twocolumn,floatfix,superscriptaddress,nofootinbib,groupedaddress,10pt]{revtex4-1}

\usepackage{amsmath}
\usepackage{amsfonts}
\usepackage{amssymb}
\usepackage{graphicx}
\usepackage{color}
\usepackage{bm}
\usepackage{etoolbox}

\begin{document}

\title{Burstiness and fractional diffusion on complex networks}

\author{S. de Nigris}
\email{denigris.sarah@gmail.com}
\affiliation{naXys, University of Namur, Rempart de la Vierge 8, 5000 Namur, Belgium}

\author{A. Hastir} 
\affiliation{naXys, University of Namur, Rempart de la Vierge 8, 5000 Namur, Belgium}
\author{R. Lambiotte}
\affiliation{naXys, University of Namur, Rempart de la Vierge 8, 5000 Namur, Belgium}
 
 \begin{abstract}
Many dynamical processes on real world networks display complex temporal patterns as, for instance, a fat-tailed distribution of inter-events times, leading to heterogeneous waiting times between events. In this work, we focus on distributions whose average inter-event time diverges, and study its impact on the dynamics of random walkers on networks.
The process can naturally be described, in the long time limit, in terms of Riemann-Liouville fractional derivatives.  We show that all the dynamical modes possess, in the asymptotic regime, the same power law relaxation, which implies that the dynamics does not exhibit time-scale separation between modes, and that no mode can be neglected versus another one, even for long times. Our results are then confirmed by numerical simulations.
\end{abstract}
\makeatletter
\patchcmd{\maketitle}{\@fnsymbol}{\@alph}{}{} 
\makeatother

\maketitle

\section{Introduction}

Linear diffusive processes are often a good entry point for the analysis of the complex dynamics of networked systems. In spite of their apparent simplicity, their flexibility allows for the modelling of a diversity of processes, including information and epidemic spreading \citep{Barrat12}.  As an additional assumption, it is also often assumed that the diffusion takes place on a static topology. The  dynamical properties of the system are then entirely determined by the spectral properties of a matrix associated to the underlying graph, for instance its Laplacian (for random walks) or its adjacency matrix (for epidemic spreading). In this  
work, the process can also be used to explore the structure of networks from a dynamical point of view, as exploited in several algorithms for, e.g. community detection or node centrality \citep{Lambiotte2014}.

Despite its many successes, this simplified model of real-world systems is also known to present several limitations \cite{Holme12}. In particular, the implicit assumption that  events take place at a constant rate, as the realization of a stationary Poisson process, has been contradicted in a  variety of systems, in terms of  temporal correlations ~\cite{Scholtes2014, Rosvall2014}, non-stationarity~\cite{Rocha13, Horvath14,Malmgren08} but also the presence of fat-tailed distributions of inter-event times~\cite{Eckmann04, Barabasi05, Rocha10,Starnini12,karsai2011small}. The main purpose of this work is to focus on this latter aspect, and to investigate the impact of the shape of the corresponding distributions on diffusive processes. This question has a long tradition in statistical physics, in situations when the underlying network is a lattice \cite{Klafter11}. In recent years, researchers have also been interested in random structures~\cite{Vazquez07, Iribarren09, Min11, Rocha13, Jo14}, seen as models of complex networks. It is only more recently that  complex structural and temporal patterns  have been combined in a single framework \cite{Delvenne15,Speidel,Hoffmann12}. The main purpose of this work is to go one step further along this line and to consider extreme situations where the waiting time distribution has a power-law tail with diverging average waiting times.

Diffusion with diverging average waiting times is well-known to exhibit anomalous diffusion, with the variance of the process growing slower than linearly with time, when the underlying network is a regular lattice. The main novelty here is to consider a finite network with arbitrary structure, as the variance of a position is not well-defined then, and the dynamics has instead to be considered in terms of the relaxation of eigen-modes. The divergence of the average waiting time naturally leads to the formulation of the problem in terms of fractional derivatives, which already have a long tradition in the study of diffusion on regular lattices and continuous media \cite{Klafter_report00}. It is interesting to note that,
despite the frequent consideration of power-law distributions and diverging moments in networks,  the use of fractional derivatives is still mainly unexplored in this field of research. Rare exceptions include the recent work of Georgiou et al. \cite{Georgiou15} on networks with random link activation and deletion, with  Mittag-Leffler interevent-time distributions and of Riascos and Mateos on the emergence of L\'{e}vy flight processes on networks through a fractional calculus approach \cite{Riascos15}. 

This paper is organised as follows. In section \ref{sub:ctrw}, we introduce the model and describe the properties of its master equation. 
In section \ref{sub:fractional dynamics}, we show that the temporal evolution can be asymptotically described by equations whose temporal derivative is fractional, which reflects the non-Markovian nature of the process. The equations reveal that the dynamical modes asymptotically relax  in a power-law way, as expected, but also  that the contribution of all modes remains non-negligible, for arbitrary long times. Theoretical results are tested against numerical simulations on artificial networks in section \ref{sub:Simulations}. Finally, we conclude and discuss the implications of our work, in particular for the problem of model reduction, in section \ref{sec:Conclusion}.

\section{From random walks to fractional derivatives}

\subsection{Continuous-time with power-law waiting times} \label{sub:ctrw}

We focus on the dynamics of a random walker on a network composed of $N$ nodes. 
The process is encoded by structural and temporal factors. 
First, the $N \times N$ transition matrix $T$ determines the probability 
that a walker jumps from one node to one of its neighbours in the network. 
The transition matrix is defined as $T_{i,j}=A_{i,j}/k_j$ where $A_{i,j}$ is the adjacency matrix encoding the
connections and $k_j$ is the degree,i.e. the number of connections of node $j$.
If we consider a Markovian process, i.e. a random walk without memory, the Markov chain associated to $T$ completely 
determines the statistical properties of the walker trajectories on the network.

Considering now the temporal side, we introduce the waiting time distribution $\psi(t)$ which
determines the duration of the stay on a node before the walker performs a jump.
Thus this quantity directly affects the timings on the trajectory generated by $T$ and, as a consequence some particular choices of $\psi(t)$,
the overall asymptotic behaviour shall be altered in respect to the Markovian frame.
For the sake of simplicity, we have assumed in the following that the waiting time distribution is the same on each node. 

Combining the two ingredients, structural and temporal, the evolution of the $x_i(t)\in \mathbb{R}^N $, the probability for the walker to be at node $i$
reads:
\begin{equation}
 \pmb{x}(t)= \pmb{x_0}\int_t^{\infty}\psi(\Delta t)d\Delta t + \int_0^t \psi(t-\Delta t)T\pmb{x}(\Delta t)d\Delta t 
\label{eq:evolution}
\end{equation}
In Eq.~\ref{eq:evolution} the first term accounts for the possibility that a walker stays put in the node: this is clearer if we  
define the survival probability as
\begin{equation}
\Psi(t)= \int_t^{\infty} \psi(\tau) d\tau=1- \int_0^t \psi(\tau) d\tau,
\end{equation}
which is the probability that the waiting time on a site exceeds $t$.
The second term, on the contrary, describes the flow of probability coming from the interaction with the other nodes in the network.
The master equation of this process,  generalizing the classical Montroll-Weiss equation to the case of complex networks \cite{montroll1965random}, takes a more suitable expression in the Laplace domain
\begin{equation}
\pmb{x}(s) = \frac{1 - \psi(s)}{s} \pmb{x_0} + \psi(s) T \pmb{x}(s) 
\label{eq:laplace_evolution}
\end{equation}
where $\pmb{x}(s)$ is the Laplace transform of the probability $\pmb{x}(s)$, defined as
\begin{equation}
\pmb{x}(s) = \int_0^{\infty}\pmb{x}(t)e^{-st}dt
\end{equation}

In the case of a Poisson process, where the waiting time is exponential by definition, this set of equations takes a simple form in the real-time domain 
\begin{equation}
\partial_t \pmb{x}(t) =L\pmb{x}(t)
\end{equation}
and one recovers the usual rate equation for continuous-time random walks, driven by the normalized Laplacian operator $L=T-I$, which is equivalent, up to a change of variables, to basic models for opinion dynamics~\cite{Blondel2005}, the position of robots in  physical space~\cite{ali} and synchronisation~\cite{Strogatz2000}.

In that case, the dynamics is determined by the spectral properties of the Laplacian operator. 
Conservation of probability implies that the dominant eigenvector, which we assume to be unique without loss of generality, 
has an eigenvalue zero. To determine the speed of convergence to stationarity, a key quantity is  the second dominant eigenvalue, 
usually called the spectral gap~\cite{Chung96}, which is  related to important structural properties of the underlying network, such as the existence of bottlenecks and communities.

We can cast Eq.~\ref{eq:laplace_evolution} in the following form to make the Laplacian operator appear:
\begin{equation}
\left(\frac{1}{\psi(s)}-1\right)\pmb{x}(s)=\left(\frac{1}{\psi(s)}-1\right)\frac{1}{s}\pmb{x_0}+L\pmb{x}(s)
\label{eq:laplac-evol2}
\end{equation}
The Laplacian term, from one side, entangles the dynamics carrying the interactions between nodes but, from the other side 
it provides the natural basis through which decompose Eq.~\ref{eq:laplace_evolution} in distinct modes. Of course, since $L=T-I$ the transition matrix and the Laplacian share the same left and right eigenvectors basis and their spectra are related by $\lambda_\alpha^{(Tr)}=1-\lambda_\alpha^{(Lapl)}$.  We can thus express the probability $\pmb{x}(s)$ as a linear combination of the right eigenvectors $\pmb{v_\alpha}$
\begin{equation}
\pmb{x}(s)=\sum_\alpha a_\alpha(s)\pmb{v_\alpha}.\label{eq:representation}
\end{equation}
Injecting this representation in Eq.~\ref{eq:laplace_evolution} and projecting on the left eigenvectors basis (for the details of the calculation, see the Appendix \ref{sec:appendix}), we thus obtain a set of equations for the amplitudes of these modes $a_\alpha$
\begin{equation}
a_\alpha(s) = \frac{1 - \psi(s)}{s}a_{\rm init}^\alpha + \psi(s) \lambda_\alpha a_\alpha(s),
\label{eq:mode}
\end{equation}
where $\lambda_\alpha$ is eigenvalue of the transition matrix associated to the  $\pmb{v_\alpha}$ eigenvector $T\pmb{v_\alpha}=\lambda_\alpha\pmb{v_\alpha}$ (here and in the following $\lambda_\alpha=\lambda_\alpha^{(Tr)}$, we dropped the exponent $(Tr)$ for the sake of lightness). We obtain the formal solution
which leads to exponential relaxations for Poissonian processes since the $\psi(t)$ is exponential as well in this case. However, if  the probability distribution of waiting times takes another functional form, it can entail more general relaxations, as we will display in the next Section.

\subsection{Fractional dynamics} \label{sub:fractional dynamics}

In the present work, we shall deal with waiting time distributions possessing a power law tail for $t\rightarrow\infty$
\begin{equation}
\psi(t) \approx \frac{\beta}{\Gamma(1-\beta)} \frac{\tau^\beta}{t^{\beta+1}},
\end{equation}
for $0 < \beta < 1$ and the prefactor has been chosen to simplify further expressions.
By definition, the average waiting time is not defined but, by application of the Tauberian theorem, its Laplace transform takes the form
\begin{equation}
\psi(s) = 1 - (\tau s)^\beta
\label{eq:tauberian-power}
\end{equation}
in the limit of small $s$, which corresponds to the asymptotic regime $t\rightarrow\infty$ we are interested in. For random walks governed by such slow decaying power tails on lattices, it is known that the average number of steps exhibits a sublinear growth, as 
\begin{equation}
\langle n(t)\rangle = \frac{1}{\Gamma(1+\beta)} \frac{t^\beta}{\tau^\beta}.
\end{equation}
Therefore, the rate of events decreases with time,
in contrast with standard processes where the average waiting is defined and $\langle n(t)\rangle  \approx t$. This property tends to slow down the dynamics, leading to subdiffusion.
Using Eq.~\ref{eq:tauberian-power} in the expression for the evolution of the modes in the Laplacian domain leads to
\begin{equation}
a_\alpha(s) = \frac{(\tau s)^\beta a_{\rm init}^\alpha}{s (1 -  \lambda_\alpha (1 - (\tau s)^\beta))}
\label{eq:evolution-w-tauberian}
\end{equation}
For the zero mode of the Laplacian, which corresponds to $\lambda_1=1$ for the transition matrix, one recovers a stationary solution, as expected:
\begin{eqnarray}
a_1(s) = \frac{(\tau s)^{\beta} a_{\rm init}^1 }{s  (\tau s)^{\beta}}
\end{eqnarray}
For the rest of the spectrum, when $\lambda_\alpha\neq 1,\lambda_\alpha\neq0$, we can cast Eq.~\ref{eq:evolution-w-tauberian} in the following form
\begin{equation}
\left(K^\beta_\alpha s^{-\beta}+1\right)a_\alpha(s)=\frac{a_{\rm init}^\alpha}{\lambda_\alpha s},
\label{eq:rewriting}
\end{equation}
where $K^\beta_\alpha=\frac{1-\lambda_\alpha}{\lambda_\alpha}\tau^{-\beta}$.
In Eq.~\ref{eq:rewriting} we can recognize a term which carries the action of a fractional operator since we have
\begin{equation}
s^{-\beta}a_\alpha(s)=\mathcal{L}\left\lbrace _0D_t^{-\beta}a_\alpha(t)\right\rbrace,
\end{equation}
with $D^{-\beta}$ is the Riemann-Liouville integral operator of fractional order $\beta$ \footnote{In what follows, the bounds of integration shall always be $0$ and $t$, thus we shall omit them to have a lighter notation.}
\begin{equation}
_0D_t^{-\beta} f(t)= \frac{1}{\Gamma(\beta)} \int_0^t \frac{f(t')dt'}{(t-t')^{1- \beta}}.
\end{equation}
We thus can interpret Eq.~\ref{eq:rewriting} as the Laplace transform of the integral equation
\begin{equation}
a_\alpha(t)-\frac{a_{\rm init}^\alpha}{\lambda_\alpha}=-K_\alpha^\beta D^{-\beta}a_\alpha(t),
\label{frac-integral-eq}
\end{equation}
which corresponds, applying the fractional differential operator $D^{\beta}$ on the left, to
\begin{equation}
D^{\beta}\left[a_\alpha(t)-\frac{a_{\rm init}^\alpha}{\lambda_\alpha}\right]=-K_\alpha^\beta a_\alpha(t).
\label{eq:frac-diff-eq}
\end{equation}
The solution to this equation is a Mittag-Leffler function 
\begin{equation}
a_\alpha(t)=\frac{a_{\rm init}^\alpha}{\lambda_\alpha}E_{\beta}\left[-\left(k_\alpha t\right)^{\beta}\right],
\label{eq:solution-mittag}
\end{equation}
where $k_\alpha=\left(K_\alpha^\beta\right)^{1/\beta}$ and the function is defined as 
\begin{equation}
E_{\beta}\left[-\left(k_\alpha t\right)^{\beta}\right]=\sum_{n=0}^{\infty}(-1)^n \frac{(k_\alpha t)^{n\beta}}{\Gamma(\beta n+1)}.
\end{equation}
By its definition, the Mittag-Leffler function appears as a generalization of the exponential and the relaxation given by Eq.~\ref{eq:solution-mittag}
approaches an exponential one in the limit $\alpha\rightarrow1$. The interesting feature of these functions, in the scope of our analysis, is their asymptotic behaviour for $t\rightarrow\infty$ which transfers the power law relaxation, proper of the waiting time distribution, to the mode themselves. Indeed, in this limit, it holds \cite{Mainardi14}
\begin{equation}
E_\beta(-t^\beta)\sim \frac{t^{-\beta}}{\Gamma(1-\beta)},
\label{eq:approx-mittag}
\end{equation} 
therefore, in this limit, Eq.~\ref{eq:solution-mittag} can be approximated by
\begin{equation}
a_\alpha(t)=\frac{a_{\rm init}^\alpha}{\lambda_\alpha}\frac{\left(k_\alpha t\right)^{-\beta}}{\Gamma(1-\beta)}=\frac{a_{\rm init}^\alpha\tau^\beta}
{1-\lambda_\alpha}\frac{t^{-\beta}}{\Gamma(1-\beta)}
\label{eq:modes-asymptotic}
\end{equation}

As for the case $\lambda_\alpha=0$, by substitution in Eq.~\ref{eq:evolution-w-tauberian} we obtain
\begin{equation}
a_0(s)=\tau^\beta a_{\rm init}^0 s^{\beta-1},
\label{eq:rogue-mode}
\end{equation}
which can be recast in
\begin{equation}
s^{-\beta}a_0(s)=s^{-1}\tau^\beta a_{\rm init}^0.
\end{equation}
With the same arguments than before, we recognize the Laplacian transform of the equation
\begin{equation}
\mathcal{L}\left\lbrace D^{-\beta}a_0(t)\right\rbrace=\mathcal{L}\left\lbrace a_{\rm init}^0\tau^\beta\right\rbrace
\end{equation}

Thus the modes associated to the eigenvalue $\lambda=0$ evolve, inverting the Laplace transform, as
\begin{equation}
a_0(t)=\frac{\tau^\beta a_{\rm init}^0}{\Gamma(1-\beta)}t^{-\beta}
\label{eq:mode zero}
\end{equation}
We can now, from Eqs.~\ref{eq:modes-asymptotic}-\ref{eq:mode zero}, reconstruct the asymptotic behaviour of the 
walker probability to be at a given node:
\begin{equation}
 \pmb{x}(t)= a_{\rm init}^1\pmb{v_1}+\sum_{\alpha=0,\\ \alpha\neq1}^{N}\frac{a_{\rm init}^\alpha\tau^\beta}
{1-\lambda_\alpha}\frac{t^{-\beta}}{\Gamma(1-\beta)} \pmb{v_\alpha}.
\label{eq:probability-evol}
\end{equation} 
To further appreciate the implications of Eq.~\ref{eq:probability-evol}, we can recast it as
\begin{equation}
\pmb{x}(t)= a_{\rm init}^1\pmb{v_1}+\frac{\tau^\beta}{\Gamma(1-\beta)t^{\beta}}\sum_{\alpha=0,\\ \alpha\neq1}^{N}\frac{a_{\rm init}^\alpha}
{1-\lambda_\alpha} \pmb{v_\alpha},
\label{eq:probability-evolution-final}
\end{equation}
where the eigenvalues $\lambda_{\alpha}$ in the sum are, of course, in the interval $\left[-1,1\right)$.

 It appears then evident that the spectrum of the transition matrix, and thus the Laplacian, does not organize any more the diffusion in a hierarchy of timescales, allowing to discriminate between groups of nodes where the diffusion has already relaxed to the stationary state and others which, in the worst case, shall relax with the slowest possible timescale, i.e. the second smaller eigenvalue of the Laplacian. In contrast, the introduction of a power law waiting time probability distribution, asymptotically induces in all the modes the same relaxation (Eq.~\ref{eq:modes-asymptotic}), thus effectively wiping out the localization effect brought by the Laplacian spectrum by itself, as we shall see in Sec.~\ref{sub:Simulations}. We note our result is coherent with the the findings of \cite{Delvenne15}, derived in situations when the waiting time distribution has an exponential tail.

Summing up the general gist of this section, we passed from the integral equation in Eq.~\ref{eq:evolution} of a continuous time random walk to a fractional differential equation  (Eq.~\ref{eq:frac-diff-eq}) in the limit of $t\rightarrow\infty$. Now, to conclude, we would like to note that this passage might be delicate one as it is not a mere change of representation. Indeed, the choice of formulation involves the correct handling of the initial conditions: as an example, in \cite{Hilfer00}, it was shown how the solution of a fractional diffusion equation can eventually turn out not being a proper probability distribution; in the present case however our starting point, Eq.~\ref{eq:evolution}, was shown to be correctly related to a CTRW process. Another possible caveat comes from the intrinsic non-Markovian character embedded by the fractional time derivative: as the system keeps track its whole history, in \cite{Hilfer03}, the author considers two waiting time distribution, differing by a fast decaying term, that can lead to two different diffusion equations in the \emph{continuum limit}; nevertheless it must be stressed that, in the context of networks, the Laplacian operator always remains discrete when passing from a random walk description to a diffusive one and, therefore, such incongruence do not arise.  
\subsection{Simulations} \label{sub:Simulations}
\begin{figure}[!h]
\textbf{a}\includegraphics[width=8cm,trim={9mm 0 0 0},clip]{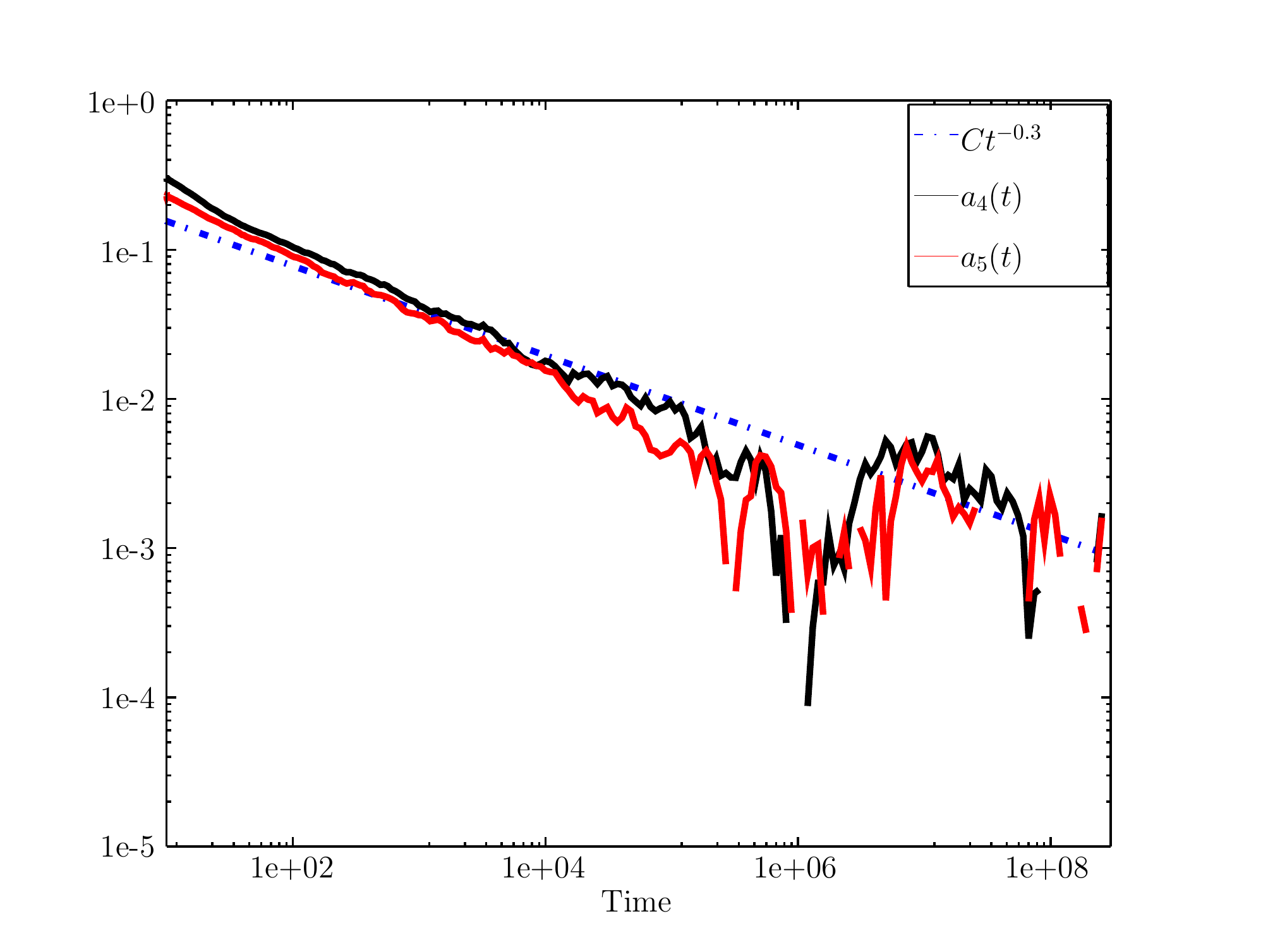}\\
\textbf{b}\includegraphics[width=8cm,trim={9mm 0 0 0},clip]{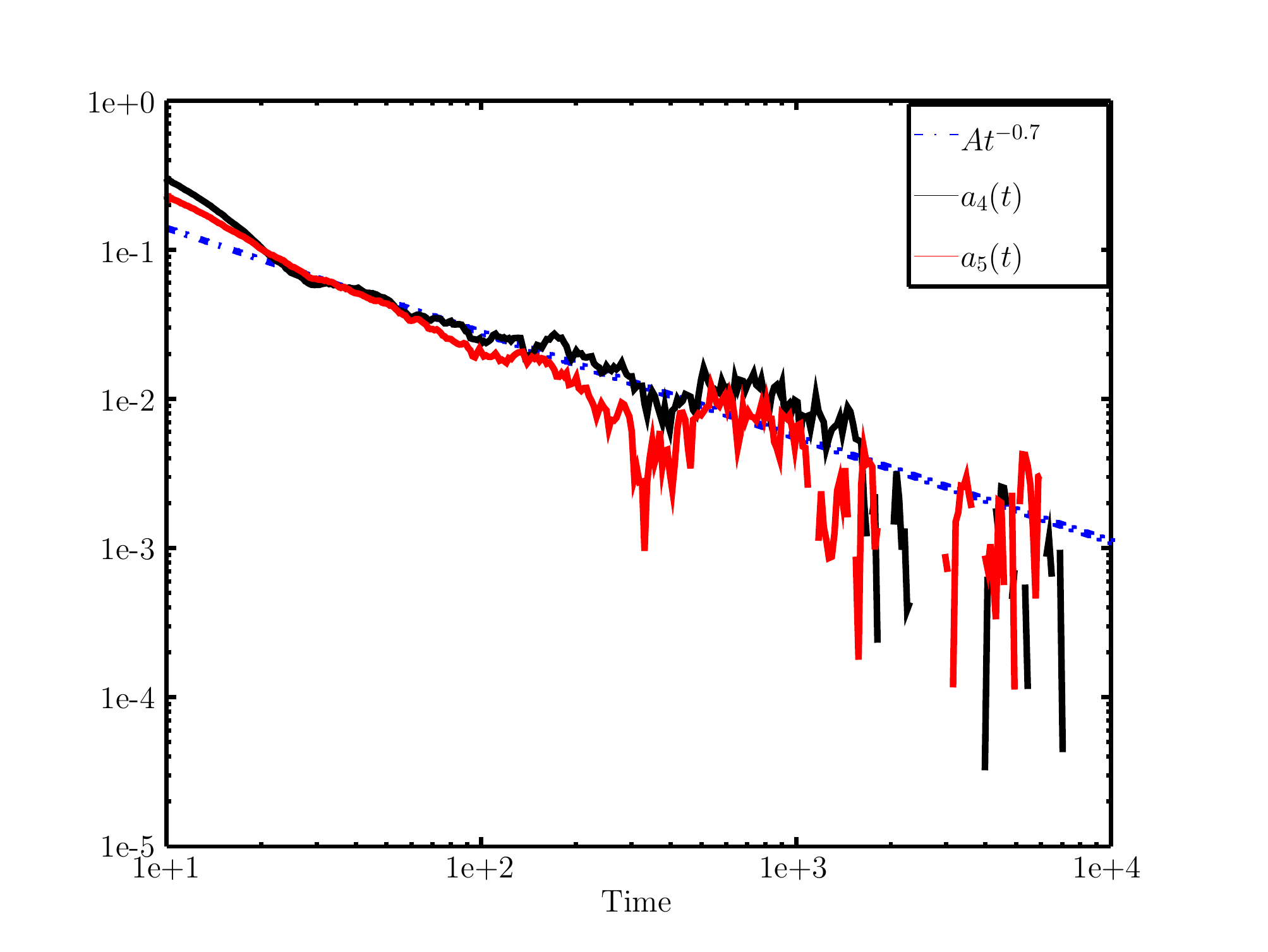}
\caption{\label{fig:relaxmode} Logarithmic plot of the $a_4(t)$ and $a_5(t)$  mode associated respectively to $\lambda_4=-0.6876$ and $\lambda_5=-0.4050$ for the transition matrix of the simple network in Fig.~\ref{fig:network} for $10^5$ walkers and (a) $\beta=0.3$ and (b) $\beta=0.7$.}
\end{figure}

In this Section, we turned to simulations in order to
get a grip on the analytical results we have shown in the previous section.
We thus explored two paths: we first considered a simulation meant to reproduce the random walk biased with a power law waiting time distribution. Through the simulation
we reconstructed the evolution of the probability $\pmb{x}(t)$ looking at the occupation of the different nodes by $N$ walkers. The latter performed or not
an hop to an adjacent node at times drawn from a waiting time distribution with various exponents $\beta$ (Figs.\ref{fig:relaxmode}a-b).
\begin{figure}
\includegraphics[width=5cm]{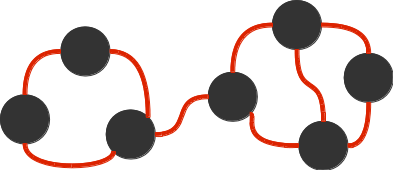}
\caption{Toy network used as a substrate for the random walks of Figs.~\ref{fig:relaxmode}.}
\label{fig:network}
\end{figure} As displayed in Figs.~\ref{fig:relaxmode} we indeed observe a slow power law relaxation over several orders of magnitude in contrast with the fast exponential relaxation that arises when $\psi(t)$ is exponential as well.

\begin{figure}
\textbf{a}\includegraphics[width=3.5cm]{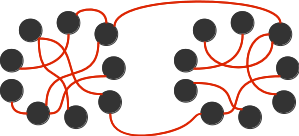}~~~
\textbf{b}\includegraphics[width=3.5cm]{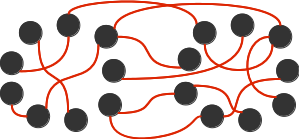}
\caption{Examples of networks used for the solutions in Fig.~\ref{fig:relaxmodes_approx}: (a) for small $p$ and (b) for $p\rightarrow1$.}
\label{fig:network_module}
\end{figure}
\begin{figure}
~~~\textbf{a}\includegraphics[width=\columnwidth,trim={9mm 0 0 0},clip,width=1.09\columnwidth]{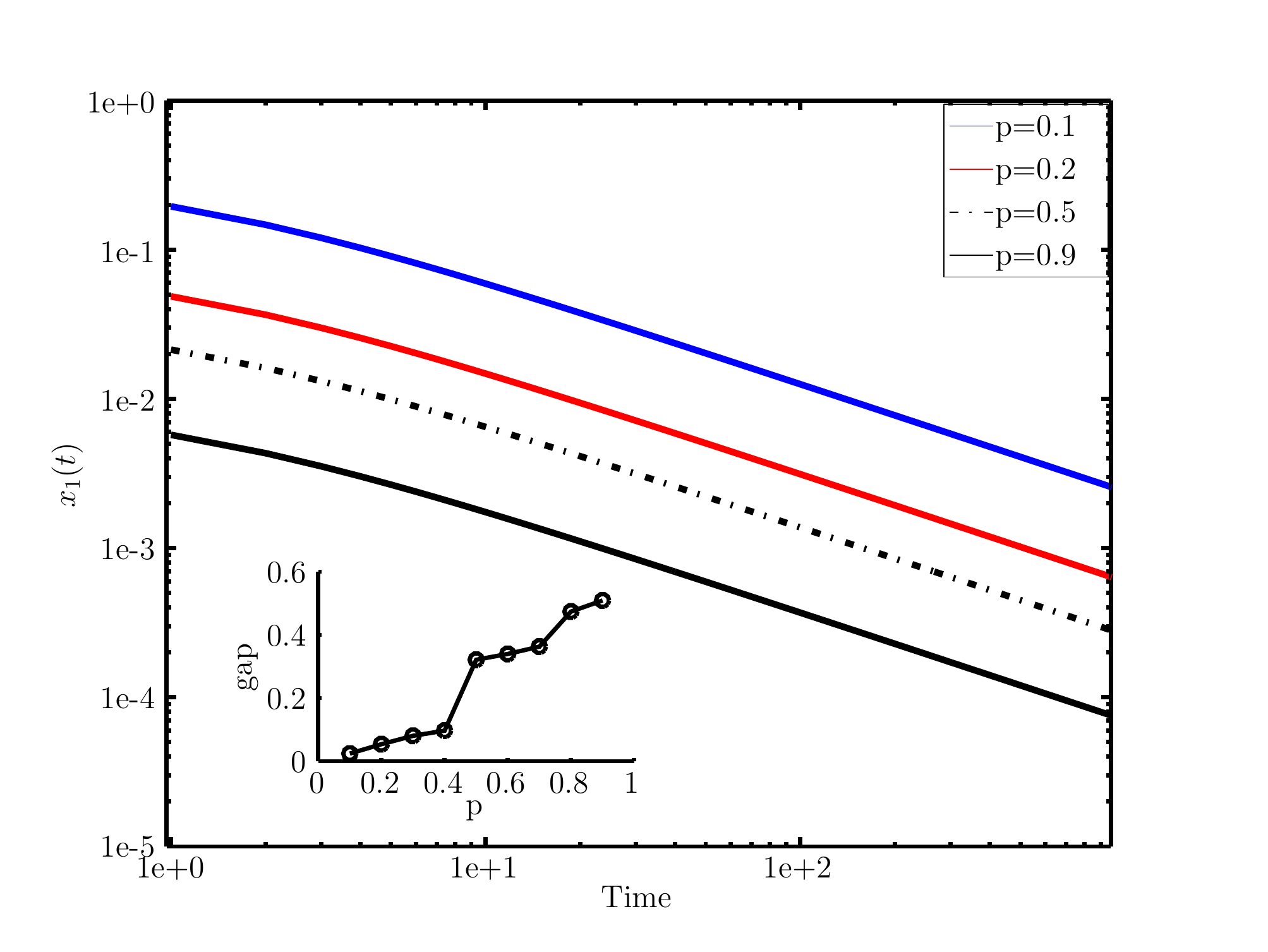}\\
\textbf{b}\includegraphics[width=\columnwidth,trim={5mm 0 0 0},width=1.08\columnwidth,clip]{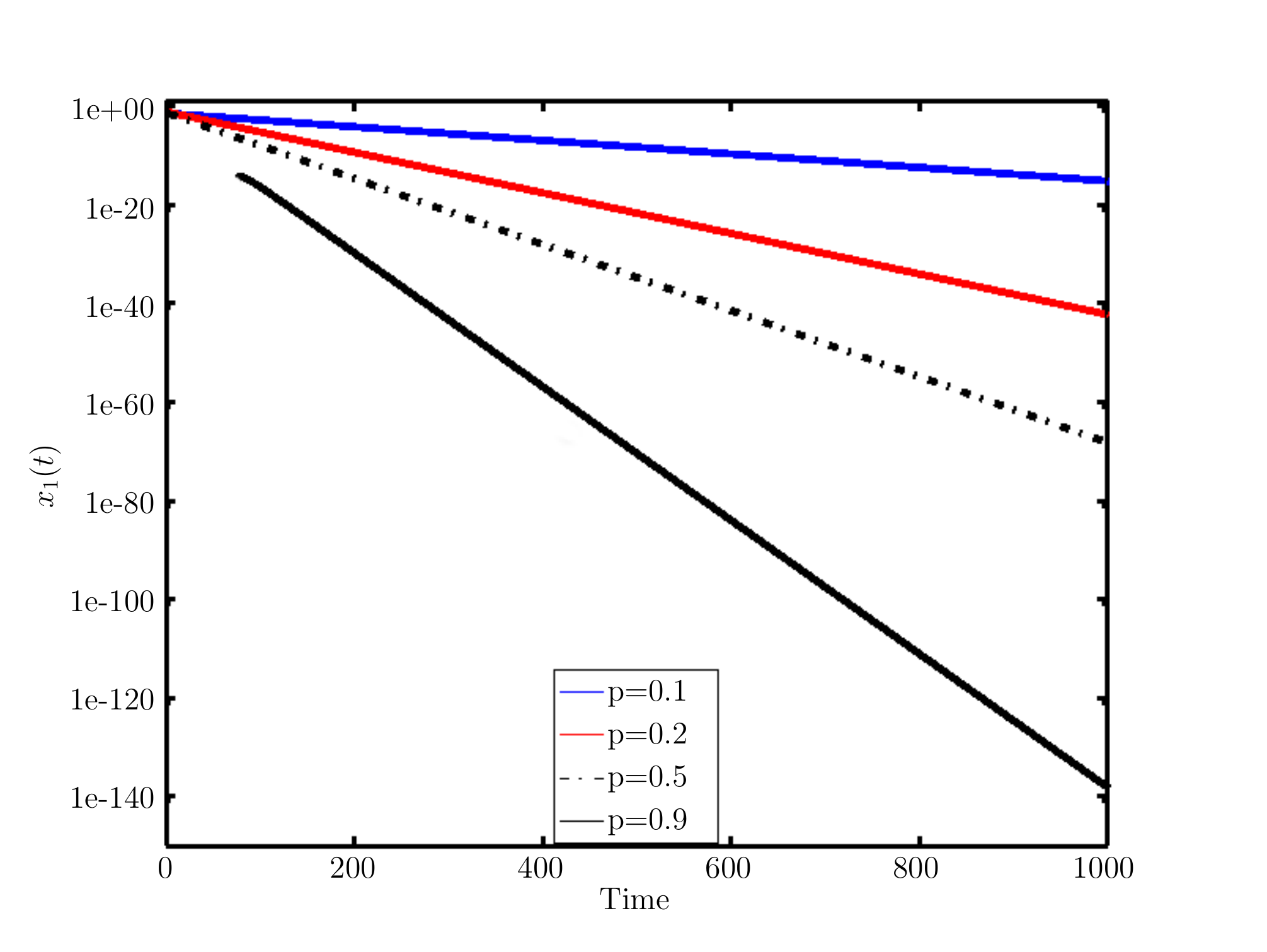}
\caption{(a) Relaxation towards equilibrium of the probability distribution $x_i$ at a node for different probabilities $p$ of connection between the two modules in the case of a power-law waiting time distribution, here $N=10$ for each module and $\beta=0.7$.
(inset) Variation of the spectral gap for the Laplacian matrix spectrum $\lambda^L=1-\lambda^T$ in correspondence to the bridging probability $p$. (b) Probability distribution $x_1(t)$ when $\psi(t)\propto \exp(-t/\tau)$ for the same networks of (a). Different eigenvalues now correspond to different relaxation times, and modes associated to small eigenvalues tend exhibit a slower relaxation, therefore dominating for long times.\label{fig:relaxmodes_approx}}
\end{figure}

We then performed a second check, which is now meant to show some implications of the coherent modes relaxation in Eq.~\ref{eq:probability-evolution-final}:  we took into account a network composed by two modules of $N=10$ each that we can, according to some probability $p$, progressively connect (Fig.~\ref{fig:network_module}). Therefore for $p\rightarrow1$ the initial two-modules structure gets blurred and finally, there is no structural signature evidencing modularity.
From the spectral point of view, this transition towards non-modularity is signalled by the spectral gap, i.e. the second smaller of the Laplacian matrix: thus, the more it approaches zero, the slower diffusion reaches the equilibrium. Therefore, in a "classical" setting in which hops are exponentially distributed there is a clear fingerprint in the dynamics entailed by a structural property. Now, in our case of study, the temporal aspect which accounts for the intermittent nature of the links, dominates over the structural one: in Fig.~\ref{fig:relaxmodes_approx}a we display the solutions for the asimptotic regime in Eq.~\ref{eq:probability-evolution-final} where, increasing the $p$ parameter we progressively erase the modular structure and correspondingly the spectral gap increases, as displayed in the inset of Fig.~\ref{fig:relaxmodes_approx}a. 
This latter tendency of the spectral gap would give a faster relaxation to equilibrium, as in Fig.~\ref{fig:relaxmodes_approx}b, but, in our case of study, this effect is hidden by the complex temporal behaviours: in Eq.~\ref{eq:probability-evolution-final}, we see that only two factors play a role in the relaxation, namely the initial conditions and the exponent $\beta$ and, indeed, the probability distribution in Fig.~\ref{fig:relaxmodes_approx}a relaxes with the same speed on structures characterized by different spectral gaps.

\section{Conclusion} \label{sec:Conclusion}

In this work, we have focused on the asymptotic properties of random walk processes taking place on networks, where the distribution of  waiting times between events presents a power-law tail associated to a diverging average.
 As  shown in Sec.~\ref{sub:ctrw}, this process can be described by a generalized master equation (Eq.~\ref{eq:evolution})  determining the time evolution of the probability for the walker to be in a node $\pmb{x}(t)$. We then expressed this vector in the basis formed by the eigenvectors of the transition matrix and considered the problem in the Laplace domain, in order to derive simple algebraic equations for the relaxation of the eigenmodes of the process. As a next step, we considered  the asymptotic limit $t\rightarrow \infty$, or equivalently $s \rightarrow 0$, where the dynamics can be rewritten in terms of 
 Riemann-Liouville fractional operators. We first showed that all the non-stationary modes exhibit the same asymptotic power-law relaxation, independently of their eigenvalue, in contrast with the standard exponential relaxation taking place for Poisson processes. In this classical case, modes do not exhibit the same relaxation, but one observes instead  a hierarchy of relaxation time-scales determined by their  eigenvalue. The existence of different time scales is crucial in order to define reduced models where fast scales can be neglected in front of slower ones, but also to uncover interesting structures from the corresponding eigenvectors. In our case, however, the extreme fluctuations of the waiting times hinder the importance of the underlying  structure, and imply that the amplitudes of the different eigenmodes remain of the same importance, even at large times. 
As expected, the power law tail of the $\psi(t)$ also significantly slows down the relaxations, as we confirmed by means of numerical simulations. 
  
\appendix

\section{Projection on the eigenmodes}\label{sec:appendix}
The transition matrix $T_{i,j}=A_{ij}/k_j$ does not possess a basis of orthonormal eigenvectors due to its asymmetry. Nevertheless, its right and left eigenvector basis,
$\left\lbrace\pmb{v_a}\right\rbrace$ and $\left\lbrace\bf{u_a}\right\rbrace$ respectively, can be related with the orthonormal eigenvector basis $\left\lbrace\bf{n_a}\right\rbrace$ of the symmetric matrix defined as
$N_{ij}=A_{ij}/\sqrt{k_i k_j}$ which is related to the normalized Laplacian \cite{Chung96}.
By definition, for $u_a$ it holds $\sum_i u_a^i T_{i,j}=\lambda_a u^j_a$ which can be rewritten, using the expression for $T_{i,j}$, as
\begin{equation}
 \sum_i u_a^i \frac{\sqrt{k_i}}{\sqrt{k_j}}\frac{A_{ij}}{\sqrt{k_i k_j}}=\frac{1}{\sqrt{k_j}}\sum_i u_a^i\sqrt{k_i}N_{i,j}= \lambda_a u^j_a,
\end{equation}
therefore the eigenvectors of the normalized laplacian can be expressed by the left eigevectors of $T_{i,j}$ through $n_a^j=\sqrt{k_j}u_a^j$.
With the same reasoning one obtains the equivalent relation for the right eigenvectors $n_a^j =  v_a^j/\sqrt{k_j}$.
The usefulness of those relations appears in the attempt to separate the modes $a_\alpha(t)$: if we consider Eq.~\ref{eq:laplace_evolution} and we inject the expansions using  the right eigenvectors in Eq.~\ref{eq:representation} we obtain

\begin{multline}
x_i(s)=\sum_\alpha a_\alpha(s)\sqrt{k_i}n_\alpha^i=\frac{1-\psi(s)}{s}\sum_\alpha a^\alpha_{\rm{init}}\sqrt{k_i}n_\alpha^i +\\ \psi(s)\sum_\alpha a_\alpha(s) \lambda_\alpha \sqrt{k_i}n_\alpha^i,
\end{multline}
where we have expressed the right eigenvectors by the means of the normalized laplacian ones $\left\lbrace\bf{ n_\alpha}\right\rbrace$.
Now, in order to isolate the different modes, we project the probability vector $\bf{x}$ on the left eigenvectors basis $\left\lbrace\bf{ u_\alpha}\right\rbrace$ using, again, their relation with the $\left\lbrace\bf{ n_\alpha}\right\rbrace$
\begin{multline}
\sum_i \frac{n_{\beta}^i}{\sqrt{k_i}} x_i(s) = \sum_\alpha a_\alpha(s) \sum_i \frac{n_{\beta}^i}{\sqrt{k_i}}\sqrt{k_i}n_\alpha^i=a_\beta(s)=\\
\frac{1-\psi(s)}{s}\sum_\alpha a^\alpha_{\rm{init}} \sum_i \frac{n_{\beta}^i}{\sqrt{k_i}}\sqrt{k_i} n_\alpha^i+\\\psi(s)\sum_\alpha a_\alpha(s) \lambda_\alpha \sum_i \frac{n_{\beta}^i}{\sqrt{k_i}}\sqrt{k_i}n_\alpha^i.
\end{multline}
Since $\sum_i n_{\beta}^i n_\alpha^i=\delta_{\alpha,\beta}$ the previous equation gives 
\begin{equation}
a_\beta(s) = \frac{1 - \psi(s)}{s}a_{\rm init}^\beta + \psi(s) \lambda_\beta a_\beta(s),
\end{equation}
which is our Eq.~\ref{eq:mode}.


\begin{thebibliography}{63}
\providecommand{\natexlab}[1]{#1}
\providecommand{\url}[1]{\texttt{#1}}
\expandafter\ifx\csname urlstyle\endcsname\relax
  \providecommand{\doi}[1]{doi: #1}\else
  \providecommand{\doi}{doi: \begingroup \urlstyle{rm}\Url}\fi
\bibliographystyle{plain}


\bibitem[Barrat et~al.(2012)Barrat, Barth\'elemy, and Vespignani]{Barrat12}
Barrat, A., Barth\'elemy, M. \& Vespignani, A.
\newblock \emph{Dynamical {P}rocesses on {C}omplex {N}etworks.}
\newblock Cambridge University Press, Cambridge (2012).

\bibitem{Lambiotte2014}
R Lambiotte, JC Delvenne, M Barahona, 
Random Walks, Markov Processes and the Multiscale Modular Organization of Complex Networks, 
Network Science and Engineering, IEEE Transactions on 1 (2), 76-90 (2014)

\bibitem[Holme and Saram\"aki(2012)]{Holme12}
Holme, P. \& Saram\"aki, J.
\newblock Temporal networks.
\newblock \emph{Phys. Rep.} { \bf 519}, 97--125 (2012).


\bibitem[Rosvall et~al.(2014)Rosvall, Esquivel, Lancichinetti, West, and Lambiotte]{Rosvall2014}
Rosvall, M., Esquivel, A.V., Lancichinetti, A., West, J.D. \& Lambiotte, R.
\newblock Memory in network flows and its effects on spreading dynamics and community detection.
\newblock \emph{Nature Comm.} { \bf 5} (2014).


\bibitem[Scholtes(2014)]{Scholtes2014}
Scholtes, I. et~al.
\newblock Causality-driven slow-down and speed-up of diffusion in non-{M}arkovian temporal networks.
\newblock \emph{Nature Comm.} { \bf 5} (2014).


\bibitem[Rocha and Blondel(2013)]{Rocha13}
Rocha, L.E.C. \& Blondel, V.~D.
\newblock Bursts of vertex activation and epidemics in evolving networks.
\newblock \emph{PLoS Comput. Biol.} { \bf 9}, 3, e1002974 (2013).

\bibitem[Horv\'ath and Kert\'esz(2014)]{Horvath14}
Horv\'ath, D.X. \& Kert\'esz, J.
\newblock Spreading dynamics on networks: the role of burstiness, topology and non-stationarity.
\newblock \emph{New J. Phys.} { \bf 16}, 7, 073037 (2014).

\bibitem[Malmgren et~al.(2008)Malmgren, Stouffer, Motter, and Amaral]{Malmgren08}
Malmgren, R.D., Stouffer, D.B., Motter, A.E. \& Amaral, L.A.N.
\newblock A {P}oissonian explanation for heavy tails in e-mail communication.
\newblock \emph{Proc. Nat. Acad. Sci.} { \bf 105}, 47, 18153--18158 (2008).

\bibitem[Barab\'asi(2005)]{Barabasi05}
Barab\'asi, A.-L.
\newblock The origin of bursts and heavy tails in human dynamics.
\newblock \emph{Nature} { \bf 435}, 207--211 (2005).



\bibitem[Eckmann et~al.(2004)Eckmann, Moses, and Sergi]{Eckmann04}
Eckmann, J.-P., Moses, E. \& Sergi, D.
\newblock Entropy of dialogues creates coherent structures in e-mail traffic.
\newblock \emph{Proc. Nat. Acad. Sci.} { \bf 101}, 14333--14337 (2004).

\bibitem[Rocha et~al.(2010)Rocha, Liljeros, and ]{Rocha10}
Rocha, L.E.C., Liljeros, F. \& Holme, P.
\newblock Information dynamics shape the sexual networks of {I}nternet-mediated prostitution.
\newblock \emph{Proc. Nat. Acad. Sci.} { \bf 107}, 13, 5706--5711 (2010).
\bibitem[Starnini et~al.(2012)Starnini, Baronchelli, Barrat, and
  Pastor-Satorras]{Starnini12}
Starnini, M., Baronchelli, A., Barrat, A. \& Pastor-Satorras, R.
\newblock Random walks on temporal networks.
\newblock \emph{Phys. Rev. E} { \bf 85}, 5, 056115 (2012).

\bibitem[Karsai(2011)]{karsai2011small}
Karsai, M. et~al.
\newblock Small but slow world: {H}ow network topology and burstiness slow down spreading.
\newblock \emph{Phys. Rev. E} { \bf 83}, 2, 025102 (2011).



\bibitem[Klafter and Sokolov(2011)]{Klafter11}
Klafter, J. \& Sokolov, I.M.
\newblock \emph{First {S}teps in {R}andom {W}alks: {F}rom {T}ools to {A}pplications.}
\newblock Oxford University Press, Oxford (2011).
\bibitem[Vazquez et~al.(2007)Vazquez, R\'acz, Luk\'acs, and
  Barab\'asi]{Vazquez07}
Vazquez, A., R\'acz, B., Luk\'acs, A. \& Barab\'asi, A.-L.
\newblock Impact of non-{P}oissonian activity patterns on spreading processes.
\newblock \emph{Phys. Rev. Lett.} { \bf 98}, 15, 158702 (2007)

\bibitem[Iribarren and Moro(2009)]{Iribarren09}
Iribarren, J.L. \& Moro, E.
\newblock Impact of human activity patterns on the dynamics of information  diffusion.
\newblock \emph{Phys Rev. Lett.} { \bf 103}, 3, 038702 (2009).

\bibitem[Jo et~al.(2014)Jo, Perotti, Kaski, and Kert\'esz]{Jo14}
Jo, H.-H., Perotti, J.I., Kaski, K. \& Kert\'esz, J.
\newblock Analytically solvable model of spreading dynamics with non-{P}oissonian.
\newblock \emph{Phys. Rev. X} { \bf 4}, 011041 (2014).

\bibitem[Min et~al.(2011)Min, Goh, and Vazquez]{Min11}
Min, B., Goh, K.-I. \& Vazquez, A.
\newblock Spreading dynamics following bursty human activity patterns.
\newblock \emph{Phys. Rev. E} { \bf 83}, 3, 036102 (2011).

\bibitem{Delvenne15}
Delvenne, J.-C., Lambiotte, R. \& Rocha, L. E. C.. 
\newblock Diffusion on networked systems is a question of time or structure.
\newblock \emph{Nat. Comm.} {\bf 6}, 7366, 1-8 (2015).

\bibitem[Speidel et~al.(2015)Speidel, Lambiotte, Aihara, and Masuda]{Speidel}
Speidel, L., Lambiotte, R., Aihara, K. \& Masuda, N.
\newblock Steady state and mean recurrence time for random walks on stochastic temporal networks.
\newblock \emph{Phys. Rev. E} { \bf 91}, 1, 012806 (2015).


\bibitem[Hoffmann et~al.(2012)Hoffmann, Porter, and Lambiotte]{Hoffmann12}
Hoffmann, T., Porter, M.A. \& Lambiotte, R.
\newblock Generalized master equations for non-{P}oisson dynamics on networks.
\newblock \emph{Phys. Rev. E} { \bf 86}, 4, 046102 (2012).



\bibitem{Klafter_report00}
Metzler R., \& Klafter J.
\newblock The random walk's guide to anomalous diffusion: a fractional dynamics approach.
\newblock \emph{Phys. Rep., } {\bf 339}, 1-77 (2000).

\bibitem{Georgiou15}
Georgiou N., Kiss I.Z. \& Scalas E.
\newblock Solvable non-Markovian dynamic network
\newblock \emph{Phys. Rev. E } { \bf 92}, 042801 (2015).

\bibitem{Riascos15}
Riascos A.P. \& Mateos J.L.
\newblock Fractional dynamics on networks: Emergence of anomalous diffusion and L\'{e}vy flights
\newblock \emph{Phys. Rev. E } { \bf 90}, 032809 (2015).


\bibitem[Montroll and Weiss(1965)]{montroll1965random}
Montroll, E.W. \& Weiss, G.H.
\newblock Random walks on lattices II.
\newblock \emph{J. Math. Phys.} { \bf 6}, 2, 167--181 (1965).



\bibitem[Blondel et~al.(2005)Blondel, Hendrickx, Olshevsky, and
  J.N.]{Blondel2005}
Blondel, V.D., Hendrickx, J.M., Olshevsky, A. \& J.N., Tsitsiklis.
\newblock Convergence in multiagent coordination, consensus, and flocking.
\newblock \emph{Proc. 44th IEEE Conf. Decision Control} { \bf }, 2996--3000 (2005).


\bibitem[Jadbabaie et~al.(2003)Jadbabaie, Lin, and Morse]{ali}
Jadbabaie, A., Lin, J. \& Morse, A.S.
\newblock Coordination of groups of mobile autonomous agents using nearest
  neighbor rules.
\newblock \emph{Autom. Control, IEEE Trans.} { \bf 48}, 6, 988--1001 (2003).

\bibitem[Strogatz(2000)]{Strogatz2000}
Strogatz, S.H.
\newblock From {K}uramoto to {C}rawford: {E}xploring the onset of synchronization of in populations of coupled oscillators.
\newblock \emph{Phys. D} { \bf 143}, 1-4, 1--20 (2000).

\bibitem[Chung(1996)]{Chung96}
Chung, F.R.K.
\newblock \emph{Spectral {G}raph {T}heory.}
\newblock American Mathematical Society (1996).


\bibitem{Mainardi14}
Mainardi F.
\newblock On some properties of the Mittag-Leffler function $E_\alpha(-t^\alpha)$ completely monotone for $t > 0$ with $0 <\alpha< 1$.
\newblock \emph{Discrete and Continuous Dynamical Systems, Series B.} arXiv: math. MP/1305.0161 (2014).


\bibitem[Hilfer (2000)]{Hilfer00}
Hilfer R.,
\newblock Fractional Diffusion based on Riemann-Liouville Fractional Derivatives,
\newblock \emph{J. Phys. Chem. B }{\bf 104}, 3914, (2000).

\bibitem{Hilfer03}
Hilfer R.,
\newblock On fractional diffusion and continuous time random walks,
\newblock\emph{Physica A} {\bf 329}, 35-40 (2003).



\end{thebibliography}
\end{document}